\documentstyle [12pt]{article}
\begin{document}
\newcommand{\vm}{\vspace{0.2cm}}
\newcommand{\vl}{\vspace{0.4cm}}

\title{p-Adic  Field Theory limit of TGD is free of UV divergences }
\author{Matti Pitk\"anen\\
Torkkelinkatu 21 B 39, 00530, Helsinki, FINLAND}
\date{12.12. 1994}
\maketitle

\newpage

\begin{center}
Abstract
\end{center}

\vm

The p-adic description of Higgs mechanism in TGD framework provides
 excellent  predictions for elementary particle and hadrons masses\\
(hep-th@xxx.lanl.gov 9410058-62). The gauge group of TGD is just the
 gauge group of the  standard model so that it makes sense to study
the  p-adic counterpart of the standard model as a candidate for low
energy effective theory.  Momentum eigen states can be constructed
purely number theoretically and the infrared cutoff implied by the
finite size of the convergence cube of p-adic square root function
leads to   momentum discretization. Discretization solves ultraviolet
problems: the  number  of momentum states associated with a fixed
value of the propagator expression in the loop is integer and  has
p-adic norm not larger than one so that the contribution of loop
momentum squared with p-adic norm $p^{k}$ converges as
$p^{-2k-2}$ for boson loop.  The existence of the action exponential  forces
number theoretically the  decomposition  of action into  free and
interacting parts.  The free part is of order $O(p^0)$ and  must
vanish (and does so by equations of motion) and interaction part is
at most  of order
$O(\sqrt{p})$ p-adically.  p-Adic coupling constants are of form
$g\sqrt{p}$: their real counterparts are obtained by  canonical
identification between p-adic and real numbers. The discretized
version of Feynmann rules of real theory  should give   S-matrix
elements but Feynmann rules guarantee unitarity in formal sense
only.  The unexpected  result is the upper bound $L_p=L_0/\sqrt{p}$
($L_0\sim 10^3\sqrt{G}$) for the size of p-adic convergence cube from
the cancellation of infrared divergences so that p-adic field theory
doesn't make sense above length scale $L_p$.

\newpage

\tableofcontents
\newpage

\section{Introduction}

The description of Higgs mechanism in TGD framework provides excellent
 understanding of particle masses \cite{padmasses}.
The cornerstones of the approach are following:
\vm

\noindent a)  The existence of p-adic square root in the vicinity of
 p-adic real axis
 implies four-dimensional algebraic extension of p-adic numbers
identifiable locally as p-adic spacetime.  p-Adic version of conformal
invariance is suggested both by the criticality of  TGD:eish  Universe
at quantum level as well as   the existence of  minima of K\"ahler
action, which are p-adically analytic maps from p-adic $M^4$ to p-adic
$CP_2$ in flat space approximation
\cite{padTGD}.  The commutators of  infinitesimal conformal symmetries
 with $N=1$ supersymmetry generated by the right handed neutrino  plus
related kappa symmetry extend the  conformal invariance to super
conformal invariance. \\
 b) Super conformal invariance  together with basic assumptions
of TGD leads to a unique
 identification of elementary fermions and bosons as  tensor products of
representations of p-adic Super Virasoro algebra (Kac Moody spinors).  \\
 c)  The support of  p-adic  square root function in the vicinity of
 p-adic  real axis  can be regarded as p-adic version of light cone and
consists of convergence cubes (rather than spheres) of p-adic square
root function (see the first paper of \cite{padmasses}).  This suggest
that  the construction  of p-adic conformal field theory limit should
reduce to the construction of n-point functions for conformal field
theory defined on convergence cube of square root function.   \\ d)
p-Adic convergence cube can  be regarded as a particle like object in
length scales above the size of convergence cube  and the suggested
rough formulation of p-adic conformal field theory limit in the first
paper of \cite{padmasses} was based on this  particle concept.  The mass
calculations  however demonstrate that in good approximation particles
can be regarded as boundary components so that the 'points' of  n-point
functions correspond to boundary components  of p-adic convergence
cube.   In point particle limit  it does not matter whether  the
boundaries of small topologically condensed 3-surfaces  or the
boundaries of smal holes drilled on the background surface are  in
question. If this the case then conformal field theory treating
boundaries (of,  say,  holes drilled on p-adic convergence cube)  as
point like objects described by Kac Moody spinors inside p-adic
convergence cube should  provide an excellent description of particle
physics phenomena.  Even more, since Planck mass excitations are
expected to have small effect on low energy physics and the gauge group
of TGD is the gauge group of standard model the p-adic version of
standard model might provide good  approximation for the theory.

\vm

The hard part of the job is the explicit construction of the conformal
 field  theory for Kac Moody spinors  inside the convergence cube. This
requires
 the generalization of ordinary gauge field theory defined  for finite
component fields to a gauge field theory for infinite component fields
provided by Kac Moody spinors.  The main technical problem seems to be
 the construction of vertices: what is required is to find Super
Virasoro invariant action of  a state in N-S type representation on  N-S
or  Ramond type representation. A related task is to formulate general
conformal field theory limit using p-adic version of K\"ahler Dirac
action used to define configuration space metric and spinor structure
and to  show that super conformal invariance indeed results and that the
Kac- Moody spinor concept developed during the mass calculations
emerges  naturally from this formalism.  The work related to both these
problems is in progress.

\vm

In this paper a more modest approach is adopted.   p-Adic thermodynamics
predicts low energy mass spectrum with excellent accuracy and gauge
group is just standard model gauge group.   Therefore a good guess is
that p-adic YM theory more or less  identical with standard model
(without Higgs field)  should provide a good approximation of p-adic
conformal field theory   at non Planckian energies. In this paper the
general conceptual framework necessary for the construction and physical
interpretation of the theory is  studied.  \\ i) The relationship
between p-adic  and real unitarity and probability  concepts  makes
possible the physical interpretation of the theory.  Some strikingly new
effects are predicted (see the fifth paper of
\cite{padmasses}).\\ ii)  The construction of momentum eigenstates as
p-adic   planewaves involves   elegant  number theory and as  predicts
number theoretic mechanism for generation of new physically interesting
length scales.  \\ iii)  Discretization of momenta
 by the necessary infrared cutoff implied by number theory and finite
 size  of
convergence cube implying automatically the absence of  ultraviolet
 divergences.   \\
iv) Perturbation theory  has purely number theoretic justification.
   Free
theory
 corresponds to $O(p^0)$ sector and the existence of the action exponential
requires the vanishing of free field action. Same requirement implies that
interactions  to higher powers of $\sqrt{p}$.  In particular,  the
effective values of  p-adic gauge couplings are effectively of the form
$g\sqrt{p}$, where $g$ is rational number and their real counterpars
obtained by canonical identification are indeed what they should be.
   \\
v) At tree level the predictions of the theory does not seem to  deviate
very much from the predictions of ordinary field theory. The general
features of the coupling constant evolution, in particular breaking of QCD
perturbation  theory at length scale $L_p= \sqrt{p}L_0$   (of order
hadronic length scale) can
be understood number theoretically.  In fact,  all
elementary particles  can appear as quantum states only below the length
scale $L_p$ in accordance with the TGD result that 3-surfaces associated
 with
charged particles have finite size for topological reasons \cite{padTGD}.
Therefore field theory description using elementary particles as basic
dynamical objects doesn't seem to work above the length scale $L_p$.

\vm

The results concerning the concept of p-adic planewave and the
 absence of UV divergences are expected to generalize as such to the
 more
general gauge  theory formulated for Kac Moody spinors and also to the
p-adic conformal field theory formulated  in terms  of K\"ahler Dirac
action.

\section{p-Adic unitarity and probability concepts}

 p-Adic unitarity and probability
concepts discussed in \cite{padTGD} lead  to highly nontrivial conclusions
concerning the general structure of S-matrix.  S-matrix can be expressed as

\begin{eqnarray}
S&=& 1+i\sqrt{p} T\nonumber\\
T&=&O(p^0)
\end{eqnarray}

\noindent for $p \ mod \ 4=3$ allowing imaginary unit in its
 four-dimensional
 algebraic extension. Using the form $S= 1+iT$, $T= O(p^0)$  one
 would obtain in
general transition rates of order inverse of Planck mass and theory
 would
have
nothing to do with reality.   Unitarity requirement implies iterative
expansion
of $T$ in powers of $p$ (see fifth paper of \cite{padmasses}) and  the
 few
lowest powers of $p$ give extremely
 good
approximation for physically interesting values of $p$.

\vm

 The  relationship between  p-adic and real probabilities involves   the
hypothesis (for details see the fifth paper of \cite{padmasses}) that
transition probabilities depend on the experimental resolution.
Experimental resolution is defined by the decomposition of the  state
 space
$H$  into direct sum $H=\oplus H_i$ so that experimental situation cannot
differentiate between different states inside $H_i$.   To each resolutions
there are associated different real transition probabilities unlike in
ordinary quantum mechanics.  Physically this means that   the experimental
arrangements,  where one monitors each state in $H_i$ separately differ
 from
the situation,  when one only looks whether the state belongs to $H_i$.
One  application is related to momentum space resolution dependence of
transition probabilities. More exotic application described in the fifth
paper of \cite{padmasses} is related to  $Z^0$  decay widths: the total
annihilation rate to exotic lepton pairs (unmonitored) is essentially
zero:
 if one
would (could)  monitor each exotic lepton pair one would obtain simply sum
of the
rates to each pair.

\section{p-Adic planewaves}

The definition of p-adic momentum eigen states is a nontrivial problem.
 The point is that usual exponent function $f_P(x)= exp(iPx)$ does not
 make
 sense as
a  representation of momentum eigen state.   $f_P$  is  not periodic
function,   $f_P$  does not even converge if the norm of  $Px$ is not
smaller than one and the  orthogonalization of different momentum eigen
states is not possible.  For instance, the sum $f_P$ over discretized
argument $x$ does not  in general vanish since lowest order contribution is
just the number of points $x$.  This state of affairs suggests that
 p-adic momentum concept involves number theory.  It turns out that
this is the case and that momentum space has natural fractal structure.

\subsection{The concept of primitive root}

The fundamental  requirement  for planewave is  periodicity.   If
the size of the p-adic converge cube, assumed for simplicity to be
one-dimensional,   is $L_p=\sqrt{p}L_0$ then there exists an elegant
 manner
to define planewaves.   Ultraviolet  cutoff  at $L_0 \sim 10^3 \sqrt{G}$
implies
 discretization in x-space
and one can label the points of cube by numbers $ x= 0,.1,....,p-1$:
 it turns out that
this restriction is not in fact necessary but simplifies the  argument.
   The basic observation is that  p-adic numbers allow roots
 $a=a_0+a_1p+...$
 of unity satisfying the condition $a^n=1$ for some values of $n$. The
condition reduces in lowest
order to the condition   $a_0^n=1 \  mod \ p$.  One can interpret $a_0$  as
an element of finite field $G(p,1)$ \cite{Book}.
The obvious idea to use $p$:th
root of unity and its powers to define planewave basis containing p
 states.  $p$:th
root of unity do not exist however unless one performs extension of
p-adic
 numbers.
$p-1$:th root however exists and following facts hold true \cite{Book}. \\
 a)
$a_0^{p-1} \ mod \ p=1$  is identically satisfied for any $a_0$
in $G(p,1)$.  From
this it follows that the order  $n$ of     $a_0$ is always factor of
 $p-1$ so that
only finite number of orders  ($<p$) are possible for $a_0$ and also for
$a$. \\ b)  For the so called
 primitive roots
allowed by any prime the order is maximal: $n=p-1$.  If $m$ does not
 divide $p-1$
then also $a_0^m$ is primitive root. \\ c) The number of roots for
 arbitrary integer  $n$ is given by
$\Phi (n)$ defined as  the number of integers $k<n$ not dividing  $n$,
 $k=1$
included.  For $n=p$  one has clearly $\Phi (p)=p-1$ corresponding to
 numbers $a_0=1,....,p-1$.

\subsection{ p-Adic planewaves with momenta $k=0,...,p-1$ and number
theoretic generation of length scales  }

What comes first into mind is to define plane waves as functions

\begin{eqnarray}
f_k(x) &=& a^{kx}, k=0,1,...,p-2
\end{eqnarray}

\noindent  where
 $a$ is some p-adic primitive root of $1$ modulo $p$ and $k$ is an
integer
 running
from $k=0$ to $p-2$.   There are $\Phi (p)= p-1$ different plane waves
 with
this definition and this looks problematic since $p$ planewaves are
expected on physical grounds. The lacking state should obviously
correspond
to momentum $k=p-1$ and indeed does so. The point is that this state
is not
identical with  $k=0$ state  p-adically as suggested by $a^{p-1}=1$.
  This
can be seen by considering $k=p-1$ planewave at points of form
 $x=y/(p-1)$.

\vm

  The
conjugate of the p-adic planewave is just $ a^{-kx}$,  which is well
 defined
in $G(p,1)$ as well as p-adically.  The sum of $f_k(x)$ over
 $x=0,1,...p-1$
vanishes
modulo p and also p-adically.  This follows from the decomposition of
the polynomial $\sum_0^{p-1} z^k$ to product of terms $z-a^k$,  where
 $a$ is
primitive root.     This means that ortonormalization modulo $p$ is
guaranteed.
  In practice x-space  discretization
does not
matter  since p-adic field theory limit applies only at length scales
above the
cutoff scale of order $10^{3}$ times Planck length \cite{padTGD}.

\vm

  The natural identification  of the  real counterpart of  momentum
$P$ is as integer proportional to $k$: $P/2\pi =k$.  In fact, it is
 $P/2\pi$,
which appears in all formulas of p-adic Higgs mechanism rather than $P$
so
 that the p-adic nonexistence does not produce problems.
  The
real counterparts of the momenta can be defined by adding the factor of
 $2\pi$ to
the real counterpart of p-adic  momentum.   Momentum does not correspond
directly to the inverse of the  wavelength as in real context.  The
wavelength
$\lambda$  is just the order $n$ of the element $a^k$ and is a factor of
$p-1$ and  the
degeneracy associated with  a given factor $n$ is $\Phi (n)$.

\vm

One might wonder whether this selection of possible wavelengths has some
 physical consequences. The  average value  of prime divisors counted
 with
 the
degeneracy of divisor is  given by $\Omega (n)= ln(ln(n)) +1.0346$
\cite{Book}
and is suprisingly small, or order $6$  for numbers of order $M_{127}$!
 If
one can apply probabilistic arguments or \cite{Book} to the numbers  of
form $p-1$, too   then
one must conclude that very few wavelengths are possible for general
 prime
$p$!   This in turn means that to each $p$ there are associated only
 very few
characteristic length scales, which  are predictable.  Furthermore, all
 the
$p^k$-multiples of these scales are also possible if p-adic fractality
holds true in macroscopic length scales.

\vm

Mersenne primes $M_n$  can be considered as an illustrative example of
 the
phenomenon.  From \cite{Table}  one finds that  $M_{127}-1$  has  11
 distinct
prime factors
 and $3$ and
$7$ occurs three and 2 times respectively. The number of distinct length
 scales
is $3\cdot   2^{11}-1\sim 2^{12}$.  $M_{107}-1$
and  $M_{89}-1$ have    $7$ and  $11$ singly occurring
factors so that the numbers of length scales are $2^7-1=127=M_7$ and
$2^{11}-1$.
  Note that for hadrons ($M_{107}$) the number of possible wavelengths
 is
especially small:  does this have something to do with the  collective
behaviour of color confined
 quarks and
gluons?  An interesting   possibility is that length scale
generation  mechanism works even macroscopically (for p-adic length
scale
hypothesis at macroscopic length scales see \cite{padTGD}).  Long
wavelength
photons, gravitons   and neutrinos might therefore provide a completely
new
 mechanism
 for generating  periodic structures with preferred sizes of period.

 \subsection{ Fractal construction of p-adic planewaves with higher
momenta }

Particle in a box picture suggests that momentum spectrum indeed
 possesses
 infrared
cutoff but that it should be possible to realize all  momenta
$k=np^{-k}$
for $k\geq
0$.   \\
a)   Consider first momenta $k= n/p$ with p-adic norm $p$. The plane
 wave
formula can be generalized by   writing

\begin{eqnarray}
f_{p^{-r }n } (x) &=& a^ { p^{-r}nx }= f_{n}(p^{-r}x)\nonumber\\
\vert x\vert_p&\leq &p^{-r}\nonumber\\
r&=&1
\end{eqnarray}

\noindent  This function however exists for $x$ having norm not larger
  than
 $p^{-1}$ so that the state is localized and can be regarded as momentum
eigenstate in the length scale  defined by the support of the planewave,
only. \\
 b)  These planewaves are
certainly not all what is needed since the
functions representable in the basis  could have arbitrary large
gradients
only in the immediate vicinity of $x=0$.     One can
however translate the plane waves located around origin $x_a=0$ to all
 points
$x_a= n $, $n=0,p-1$, which means the replacement $x\rightarrow x-x_a$
in
the previous formula.
  In this manner one obtains  altogether $N(r=1)=p(p-1)
$  localized  planewaves with p-adic momenta $p^{-1}k$ since constant
planewave is excluded by infrared cutoff.   \\ c) The construction of
 p-adic
plane waves with p-adic momenta with p-adic norm $p^{r}$ proceeds in
 obvious
manner.   One constructs around each point

\begin{eqnarray}
x_a&=& \sum_{k=0}^{r-1} x_kp^k
\end{eqnarray}

\noindent a localized
planewave basis

\begin{eqnarray}
f_{p^{-r }k } (x-x_a)
\end{eqnarray}

\noindent  with p-adic  momenta $p^{-r}k$ and argument
$x-x_a$  and obtains in this
manner $p^{r+1}$ states.

\vm

The localization of planewaves   is not in conflict with Uncertainty
Principle  since the
localized planewaves are momentum eigenstates only in  the length scale
defined by the support of localized state: this is also clear from the
 fact
that momentum spectrum contains only the momenta $P=p^{-r}k$ but not the more
general momenta $P=\sum_r k_rp^{-r}$.  The  number of localized momentum
eigenstates with the p-adic norm of momentum not larger than $p^r$ is
$p^{r-1}(p-1)$,  where the factor $p-1$ comes from infrared cutoff
excluding
constant planewave,  and indeed equal to the possible values of momenta.
Infrared cutoff in momentum is necessary. One can obviously construct
genuine momentum eigenstates simply as products of momentum eigenstates
associated with different length scales.    If one tries to extend the
momentum range to infrared one encounters  problems with completeness of the
basis  since p-adic convergence cube does not contain  the entire
range of $x$ values for which the plane wave is well defined
(Uncertainty
 Principle!).

\vm

Momentum space has  fractal structure,
the number of momenta with p-adic norm $p^r$ being $N(r) =p^{r-1}(p-1)$   in
single spatial
 dimension.    In the limit of infinite UV cutoff the total number of
states in for dimension  $D$ is just

\begin{eqnarray}
N(tot,r)&=& p^{D(r-1)}(p-1)^D
\end{eqnarray}

\noindent  If the total number of states is defined by taking ultraviolet
cutoff to infinity the number of states is p-adically equivalent with
zero:
a  somewhat  unexpected result!   The result is important as far as
vacuum expectation values and normal ordering of oscillator operators is
considered: for instance, the normal ordering of fermion current  gives
 no
c-number term.

\vm

One can  generalize the  planewave concept somewhat. Since
the algebraic extension used allows square root one can define new planewave
basis as  square roots of the p-adic planewaves: the corresponding
momentum spectrum obviously contains half odd integers.   These planewaves
are  in general genuinely complex p-adic numbers.  The premilinary work with
conformal field theory limit suggests that the existence of two kinds of
plane waves is directly related to the existence of Ramond and N-S type
representations and that  the momentum spectrum for  Ramond/N-S type super
generators is  labeled by $Z$ integers and by $Z/2$.   Furthermore, lepton
and quark momenta should belong to $Z$  ($Z/2$) respectively.

 \section{Second quantized interacting field theory inside p-adic
 convergence cube}

In practice YM theory with standard model gauge group for leptons and
quarks plus
the some other light exotic particles predicted by the
p-adic thermodynamics should provide excellent description of the physics
 below
 non-Planck energies.  The extremely rapid convergence of the p-adic
perturbation
 theory
implies that  loop corrections coming from Planck mass excitations are
extremely
small for physical values of prime $p$ and can be neglected.  The task is
 to find whether
   it is possible to define  second quantized
 interacting  theory inside single
 convergence cube so that S-matrix has the  structure dictated by
 physicality
requirements.

\subsection{Decomposition of action into free and interacting parts number
theoretically}

The requirement that perturbation theory works requires that interaction
 terms
 $V$ are proportional to
the factor  $\sqrt{p}$ or some higher power of $p$ whereas free part of the
 action is of order $O(p^0)$ and must vanish, not only by field equations,
 but
also by the requirement that action  exponential exists p-adically.   One
 could
use functional integral formalism or  Hamiltonian approach and in both of
these approaches same number theoretic constraint is encountered.

\vm

In functional integral formalism one considers the exponent of
classical action.
      Kinetic term of  the action is of order $O(p^0)$
formally.   The exponent of
the action makes no sense unless kinetic term vanishes identically: this in
 turn
is in accordance with  equations of motion of free field theory in order
$O(p^0)$ provided action  vanishes for free field solutions.  This in
turn gives strong constraint on the action: for instance, the generation of
cosmological constant via vacuum energy density becomes impossible.
Interaction term in turn must be of order $O(\sqrt{p})$ at least so that
 S-matrix
has the required form and exponent of action exists.   Functional
integration is
over over quantum fluctuations around classical solution with vanishing
action  and one must  require that integration is only over quantum
fluctuations,  whose  contribution to the  action
of order $O(p^{1/2})$ at most.

\vm

      In Hamiltonian formulation  one  expresses the
solutions  as $M^4$ fields perturbatively
using time ordered exponential $P(exp(i\int V dt))$, where $V$ is
 $O(\sqrt{p})$
contribution of Hamiltonian.    $V$
  must be  proportional to $\sqrt{p}$ for the integral to exists.
    The existence of $P(exp(\int Vdt))$ probably
poses an upper bound for the transition time $T$ (size of the convergence
 cube
in
 time direction) since the  exponential is  not expected to exist for too
large
values of $T$.  This means that  quantum transition times are naturally
quantized.  The value of this time is naturally the duration
of time associated with single p-adic convergence cube.   There is no
 particular reason to expect that all cube sizes are possible and it turns
out that p-adic counterpart of standard model does not exist in length
scales above $L_p$.

\vm

 These   arguments suggest
 that perturbative approach  is the only manner to define p-adic quantum
theory.     The fields  are
expressible in terms of free part  $\Phi_0$ and interacting part

\begin{eqnarray}
\Phi&=& \Phi_0 +\sum_{n\geq 1}\sqrt{p}^n\Phi_n
\end{eqnarray}

\noindent   The free field $\Phi_0$   has standard expansion
 in terms of oscillator operators in one-one correspondence with
light states associated with ordinary spinors and gauge fields. $\Phi_n$
  contains off
mass shell momenta and can be solved iteratively in terms of $\Phi_0$
using the equations of motion.

\vm

One can define the oscillator operators of the
interacting theory as power series expansions and calculate S-matrix.
 Direct
manner to construct S-matrix is  LSZ reduction formula applied inside
 p-adic
convergence cube.  What one obtains  is QFT in box determined by the
 convergence
cube of square root  with infrared cutoff coming from the size of the cube.
 A reasonable  guess is that the Feynmann rules of standard gauge theory
 apply as
such since all algebraic manipulations of standard gauge theory go
 through as
such: situation is even simpler since the elimation of ultraviolet
 divergences is not needed. p-Adic unitary is guaranteed in formal sense
by Feynmann rules but the necessary infrared cutoff might lead to problems
with unitariry.

\subsection{Action and Feynmann rules}

   The ordinary YM Dirac action should describe the couplings of the
nonexotic light states.    The couplings associated with vertices containing
exotic states contain yet unknown parameters, which are predicted by p-adic
conformal field theory having as its particle content single particle states
 of second quantized theory.  The masses can be taken to be the masses
 predicted  in excellent approximation by p-adic thermodynamics.
CKM mixing
matrix  appear in quark couplings and it was found the requirement that
 topological mixing matrices are
rational  unitary matrices together with some other TGD:eish requirements
 might well determine these parameters uniquely \cite{padmasses}.

\vm

The assumption that  p-adic gauge couplings  have the  general form

\begin{eqnarray}
g_p&=& g\sqrt{p}
\end{eqnarray}

\noindent  guarantees  $V \propto \sqrt{p}$. This definition of
effective coupling  was suggested in the  fifth
paper of \cite{padmasses} and certainly provides a correct relationship
between p-adic and real coupling.
 If $g$ is rational number then the real counterpart $(g^2p)_R$
 of $g^2p$  obtained by canonical correspondence between p-adics
 and reals is  reasonably close to $g^2$ interpreted as ordinary rational
number.  In particular, for Mersenne
primes and $g^2 $, which is finite superposition of negative power of $2$
 the
$(g^2p)_R$ is numerically very near  to $g^2$ interpreted as real number.
This
implies for YM theory  that at tree graph level each internal line contains
 two
$\sqrt{p}$:s at its ends and for loop  momenta between  elementary particle
 mass
scale and Planck mass scale one $p$ in propagator so that $p$:s cancel and
one
obtains something very near to that  of ordinary gauge theory.

\vm

The naivest definition of the vertices    in perturbation
theory is not  however
the most elegant one.  One can
always redefine gauge potentials so that gauge couplings appears nowhere
except as a normalization factor  $\frac{1}{4g^2p}$ of gauge boson part of
the YM action.  This definition in turn implies that gauge boson
propagators are proportional to $g^2p$ and fermion boson vertices involve
no coupling constant.  The use of this description is
suggested also by the concept of induced gauge field, which is naturally such
that gauge couplings is included into the normalization of gauge potentials.
Of course, this definition  doesn't change the definition of,
say,  effective $\alpha_s$ but makes it easy to see that
various lowest order  loop corrections are
p-adically of same order of magnitude below $L_p$.

\vm

The definition of fermionic propagators is  not quite
straightforward since the quantity

\begin{eqnarray}
G&=&\frac{1}{p_k\gamma^k+M_{op}}\nonumber\\
M_{op}^2&=&M^2
\end{eqnarray}

\noindent involves the square root $M_{op}$ of  mass squared  of fermion
calculated in the papers \cite{padmasses} as thermal expectation value.
  Dirac
operator acts on 8-dimensional  H-spinors and must respect chirality
conservation. Therefore $M_{op}$ cannot be scalar but rather a linear
combination $a\gamma_0+ b\gamma_3$ of $CP_2$  tangent space gamma matrices
commuting with electromagnetic charge operator. $M_{op}$ is clearly the
counterpart of Higgs vacuum expectation value and the geometric
counterpart of $M_{op}$ is the $CP_2$ part of the operator $H^k\gamma_k$
appearing in the Dirac equation for induced spinors,
$H^k$ being the trace of the second fundamental form. When boundary
components are idealized to world lines one can  assume that
$H^k(CP_2)\gamma_k$ is covariantly constant and $H^k(M^4_+)\gamma_k$ vanishes
(geodesic motion in $M^4_+$).  The coefficients
$a$ and
$b$ are different for different charge states of lepton/quark so that it is
possible to find a solution to the defining  condition although the solution
need not be unique.

\vm

 The essential
difference with respect to ordinary gauge theory is  the number
 theoretic description of momentum eigenstates  and  discretization
of momenta. The discretization
solves also the problem implied by the p-adical nonexistence of   standard
momentum space measure $dV= (2\pi)^3d^3p/2E_p$ ($\pi$ does not exist
p-adically).     p-Adic discretization also   implies also the absence of
ultraviolet divergences as will be found.    The cancellation of   infrared
divergences implies that the length scale
$L_p)=L_0\sqrt{p}$ gives upper bound for the size of the p-adic convergence
cube: for larger size the sum of self energy diagrams is not p-adically
convergent.

\subsection{How to compare predictions with experiment}

The comparison of the theory with ordinary field theory and experiments is
 based on
the  concept of resolution. In momentum degrees of freedom this means that
 p-adic
momentum space is divided into cells so that different momenta inside cells
 are
not monitored experimentally  and the summation over final states is
performed
p-adically. The resulting transition amplitudes squared are mapped to reals
 by
canonical identification and after this the usual momentum space
integration
measure can be used.  In practice,   the transition rates are the
 physically
interesting quantities. In standard  QFT  the squares of  S-matrix
elements  involve square  of  momentum conserving delta function and the
rate is
obtained by dividing with the momentum  delta function interpreted as
infinite
quantization volume. In present case the quantization volume is finite
and given by the volume of the convergence cube of p-adic square root
function.

\section{Number theoretic cancellation of  UV divergences and
 necessity of infrared cutoff $L_p$ }

 The fact that momentum summation always involves integers with p-adic norm
not larger than  implies the cancellation of ultraviolet divergences
provided the imbedding of $M^4$ differs slightly from standard imbedding.
 Infrared cutoff in turn is forced by infrared
finiteness.    In the following considerations are restricted to scalar loops
but similar considerations can be applied fermionic loops and fermion self
energies.

\subsection{UV finiteness}

 Consider first the   self energy contribution,  when  bosons
propagate in the loop. The summation of all possible loop momenta can be
decomposed to sum over terms  for which the scalar function

\begin{eqnarray}
F(P_1,P,P_1-P)&=&\frac{g^2p^2}{(P^2-M_1^2)((P_1-P)^2-M^2)}
\end{eqnarray}

\noindent associated with  the loop
 is
constant.  Here we have assumed that  bosonic propagators
proportional to  $p$ to guarantee that the contributions of fermionic and
bosonic loops are of same order.  For very large loop momenta the function
is in good
 approximation
$g^2p^2/P^4$ and for a given p-adic norm $p^{-k}$ of ultraviolet momentum
squared behaves
 as
$p^{2k+2}$ p-adically.  This implies rapid convergence in ultraviolet since
the number of
 all possible
loop momenta is always integer and has p-adic norm not larger than one.
This
 means
that ultraviolet divergences are completely absent!

\subsection{The problem of ill defined degeneracy factors
and propagator poles}

UV finiteness does not yet guarantee the p-adic existence of the momentum
 sums
 associated with  self energy diagrams.    For spacelike incoming
propagator
momenta one  finds that degeneracy factor associated with certain loop
momenta is infinite. Although the p-adic norm of this numbe is not larger
than one it is ill defined.    Second problem is related to the poles of
p-adic propagator, when some of the discrete  loop momenta are on mass
shell.
  Both
problems can be circumvented by using Wick-rotation trick or by  taking
into
account the deformation of standard imbedding of $M^4$ to nonstandard but
still
flat imbedding  caused by total gravitational mass of the p-adic convergence
cube.

\subsubsection{ The problem  }

Suppose that incoming space like propagator  momentum  $P$ is given by
$(0,P,0,0)$ and denote by $(0,K,0,0)$ / $k_{\perp}$ the component of loop
momentum $k$  in direction  parallel/orthogonal to   $P$.   Strictly
speaking,
this $P$  is not allowed by infrared cutoff the situation is
essentially equivalent with that obtained assuming all components of $P$
to be
nonvanishing.  The propagator expression  associated with self energy
contribution  is for scalar particle   given by

\begin{eqnarray}
\frac{1}{(-(P-K)^2-M^2)(-K^2+k_{\perp}^2-M^2)}\nonumber \\
k_{\perp}&=&  (k_0,0,k_1,k_2)
\end{eqnarray}

\noindent  The value of propagator expression is fixed once the values of
  $k^2$ and $P\cdot k$ are fixed.

\vm

   Any lightlike vector  $k_{\perp}$ contributes to the degeneracy and there
 are infinite number of light like vectors of this kind since the equation

\begin{eqnarray}
k_0^2 &=&k_1^2+k_2^2
 \end{eqnarray}

\noindent allows infinite number of solutions.   The first solution type
 consists of vectors $(k_0,0,k_1,0)$ and $(k_0,0,0,k_1)$ and the number
solutions is $(p-1)p^N$ $N\rightarrow \infty$:  the number of solutions has
vanishing p-adic norm at the limit $N\rightarrow \infty$ so that the
contribution to self energy vanishes. In fact, this solution is excluded  by
strongest form of  infrared cutoff requiring that each momentum component is
nonvanishing.

\vm

    The second solution type consists of solutions for which $k_1$ and $k_2$
 are nonvanishing.  A short calculation shows that the general solution is of
form

\begin{eqnarray}
k_i&=& (K_0,0,K_1,K_2)\sum_{n\ge 0} \epsilon_np^{-n}\nonumber\\
K_0^2&=&K_1^2+K_2^2\nonumber\\
K_0&<&p\nonumber\\
\epsilon_n&\in& (0,1,-1)
\end{eqnarray}

\noindent where $K_i$ is integer solution of  the condition. The total
 number of  integer solutions is the number of Pythagorean triangles with
hypothenuse smaller than $p$: integers scaled  triangles are counted as
different triangles. Denoting the number of these triangles with $N(p)$ one
has
for the total degeneracy

\begin{eqnarray}
N(tot)&=& \lim_{N\rightarrow }3^N   N(p)
\end{eqnarray}

\noindent This limit  is not well defined p-adically (except in case $p=3$)
although the  p-adic  norm of limit is well defined.  If one allows only
non-negative momenta the power of $3$ is replaced with power of $2$.

\subsubsection{Wick rotation solves the problem?}

The troubles clearly  result from the possibility of light like vectors
 $k_{\perp}$ in 3-dimensional hyperplane of momentum space. Therefore
 Euclidian
signature for momentum space implies that loops sums are completely well
defined
and analytic continuation in $P$ makes it possible to define momentum sums
 for
$M$ signature.  Also poles disappear.  Wick rotation is  routinely applied
 in
ordinary QFT but the application of Wick rotation in present case is somewhat
questionable trick.

\subsubsection{ Gravitational warping solves  the problem?}

The infinite degeneracy associated with spacelike momenta  $P$ disappears if
 the standard imbedding of $M^4$ is replaced with warped imbedding, for which
some components of metric differ from their standard values by scaling.
    The phenomenon is completely
analogous  with   the existence of infinitely many imbeddings of flat
2-dimensional space  $E^2$ to $E^3$ obtained by deforming any plane of $E^3$
without stretching it.  The  physical reason for warping is gravitational
mass of the p-adic convergence cube.   The mass of the p-adic convergence
cube deforms spacetime surface and average effect is  momentum dependent
scaling of some metric components.
   In GRT gravitational warping clearly does not make any
sense. The simplest  physical consequence of the deformation is that photons
propagating in condensate spend in general longer time than vapour phase
photons
to travel a given distance \cite{TGD}.

\vm

  A rather general nonstandard imbedding of $M^4$ representing gravitational
warping is obtained by assuming that  $CP_2$ projection of spacetime surface
 is a
one-dimensional curve:

\begin{eqnarray}
s&=& KP_km^k \nonumber\\
g_{\alpha\beta}&=&  m_{\alpha\beta}-K^2 P_{\alpha}P_{\beta}
\end{eqnarray}

\noindent  Here $s$ is the curve length coordinate and    $P_k$  is total
four-momentum.

\vm

The  scaling  of the metric components has two nice consequences:\\
a) Assuming that  the spectrum of off-mass shell four-momenta is not changed
the  propagator expression has no poles since one has always
  $k^2= g^{\alpha\beta}k_{\alpha}k_{\beta}\neq M^2$
in warped metric. \\
b)    For
$m_{00}=1\rightarrow  1-\delta$  the    expression $k_{\perp}^2=0$  becomes
$k_{\perp}^2=-\delta k_0^2$ and this number becomes p-adically very large,
when
p-adic norm of $k_0$ increases. This guarantees the convergence of the
propagator expression as well as finite degeneracies for all values of $P$.

\subsection{ Necessity of infrared cutoff}

 In p-adic field theory the
dangerous region seems to be infrared rather  than ultraviolet as far as
momentum space summation is considered. Now however number theoretic    (and
physically obvious) infrared cutoff resulting from the finite size of
convergence
 comes to rescue and makes self energy finite in lowest order. Although self
energy is certainly finite the geometric sum of self energies does not
converge unless self energy is of  order $O(p)$ and therefore same order
as the momentum propagating in the loop if virtual momentum is between
particle mass and Planck mass.  This condition is not however always
satisfied.\\
 a) If the particles  propagating in the loop are massive then
all contributions to  self energy  are at most of order $O(p)$
irrespective of the size $\sqrt{p^n}L_0$ of the convergence cube.\\
 b) If
loop contains massless particles then situation changes since lowest
momentum square  in massless loop has p-adic norm $p^n$.
The point is that
at worst the  massless loop particle  gives contribution

\begin{eqnarray}
 \frac{1}{P^2} &\propto& p^{-n}
\end{eqnarray}

 \noindent to self energy.   The proportionality of boson propagators to
$p$ can give at most the power $p^k$, $k=1,2,3$ so that for sufficiently
large $n$ the self energy becomes large and the geometric sum defined by
self energy does not converge  p-adically.  One could formally define the
sum as the sum of geometric series but the resulting propagator would
contain positive power of $p$ making it extremely small so that this trick
provides only a second manner to say that something drastic happens in
length scale $L_p$.

\vm

The result means  that perturbation theory is simply not well
defined in length scales above $L_p=L_0\sqrt{p}$, which is just of order of
Compton wave length for primarily condensed  particles. \\
a) Masslessness of photon implies that
charged fermions cannot appear as states of quantum field at length scales
above $L_p$ in accordance with the basic results of classical TGD.\\
b) Masslessness of gluons in turn implies that QCD is not well defined above
$L_{107}$: a clear signal of color confinement. It is p-adic cubes with size
$L_{107}$, which become basic  dynamical units at these scales and these must
be noncolored since otherwise gluon propagators would  lead to  exactly
the same difficulty as in quark level.\\
c)   Since photon self energy at lowest order involves only massive
charged particles one might think that photon self energy is always of
order $O(p)$ so that photon progation would be possible in all length
scales. The problem is that the higher loop contributions to self energy
contains emission of arbitrary number of virtual photons from charged  loop
 particles and this makes
the self energy large in length scales larger than $L_p$.  Same applies
also to intermediate gauge bosons.

\vm

 The general conclusion seems to be that above $L_p$ quantum
field theory description at condensation level $p$ fails.   It is
however possible for a particle to suffer secondary  condensation on level
$p_1>p$ and at this level quantum field description is possible below
larger scale $L_{p_1} = \sqrt{p_1}$. In this manner one obtains condensation
hierarchy.   This fits nicely with the results of classical TGD. The
electroweak gauge fields in TGD are induced from the spinor connection of
$CP_2$ and for purely topological reasons the imbedding of charged gauge
field fails for some critical size of the 3-surface.  In particular, the
size of 3-surface associated with charged particle and in fact any particle
creating long range fields  is finite: what was called 'topological field
quantum' is formed \cite{padTGD}. It was this concept, which together with
 p-adic length
scale hypothesis led to the detailed quantitative development of the concept
 of
topological condensate  \cite{TGD,padTGD} in various length scales.
 Note that the  results obtained  make  also particle-field duality
concept very concrete.

\vm

 Of course, one can ask what replaces the QFT
description above length scale $L_p$.  One possibility is that
$L_p$ represents absolute upper size of particles at level $p$ so that this
kind of description is not needed. A second possibility is that elementary
particles are replaced with
 composite objects  (say hadrons) as fundamental fields.  In
practice, it seems the p-adic version of ordinary quantum mechanics
might provide satisfactory description above $L_p$.

\subsection{Note about coupling constant evolution}

The calculation of the details of coupling constant evolution is not easy
since the calculation necessitates the calculation of degeneracies for a
given value of propagator expression and this involves difficult number
theory.  It would not be surprising if the p-adic counterparts typical
logarithmic terms of ordinary coupling constant evolution would be
encountered.

\vm

 It is however not obvious what the p-adic counterparts of logarithms of type
$ln(Q^2/Q_0^2)$ are.  In  the fifth paper of \cite{padmasses}  it was noticed
that  logarithm defined as inverse of the ordinary exponent function exists
only for $Q^2/Q_0^2=1+O(p)$ so that each range  $Q^2=p^k (n+ O(p))$
defines its own coupling constant evolution.
 The construction of p-adic
planewaves   suggests a
more satisfactory solution of the  problem.  The correct manner to
determine
logarithm is start from the exponent  $a^x$ of p-adic primitive root and
 define
logaritm via the formula

\begin{eqnarray}
y= log_a (x)  &\leftrightarrow& x=a^y \nonumber\\
a^{p-1}&=&1, \vert x\vert_p=1, \vert y\vert \ge 1
\end{eqnarray}

\noindent Logaritm is well defined for all values of $x$ having p-adic norm
equal to one. The value of the logarithm can have arbitrary large p-adic
 norm
not smaller than one.  Therefore each power of $p$ in momentum space defines
 its
own coupling constant evolution and  coupling constant evolution
 equations in
principle involve  the initial values  $g^2(k)$ defined as

\begin{eqnarray}
g^2(k)&=& g^2(Q^2_0 = p^{-k})
\end{eqnarray}

\noindent In practice, the results about cancellation of infrared
divergences suggest that only single value of $k$ is  needed and this s
power of p involves entire energy range from the size of particle to Planck
length.

\newpage

\begin{center}
{\bf Acknowledgements\/}
\end{center}

\vm

It would not been possible to carry out the work related to p-adic TGD
without the   help of
 my friends  in concrete problems of the everyday life and I want to
express my gratitude to  them.  Also I want to thank J. Arponen,
 P. Ker\"anen,
 R. Kinnunen and J.  Maalampi,
    for practical help and interesting discussions.

\end{document}